\documentclass[usenatbib]{mn2e}
\pdfoutput=1
\usepackage[pdftex]{graphicx}
\usepackage{amssymb}

\begin{document}

\title[Over-cooled haloes at $ z \ge 10$: a route to form low-mass  first stars]{Over-cooled haloes at $ z \ge 10$: a route to form low-mass  first stars}

\author[Prieto, Jimenez \& Verde]{Joaquin Prieto$^1$\thanks{email:joaquin.prieto.brito@gmail.com}, Raul Jimenez$^{2,1,3}$, Licia Verde$^{2,1,3}$ \\
$^{1}$ICC, University of Barcelona (IEEC-UB), Marti i Franques 1, E08028, Barcelona, Spain \\
$^2$ ICREA\\
$^3$ Theory Group, Physics Department, CERN, CH-1211, Geneva 23, Switzerland
}

\maketitle

\begin{abstract}
It has been shown by \citet{ShchekinovVasiliev2006} (SV06) that HD molecules can be an important cooling agent  in  high redshift  $z\ge10$ haloes if  they undergo mergers under specific conditions so suitable shocks are created.
Here we build upon \citet{jpp3} who studied in detail the merger-generated shocks,  and show that  the conditions  for HD cooling can be studied  by combining these results with  a suite of  dark-matter only simulations.
We have performed a number of dark matter  only simulations from cosmological initial conditions inside boxes with sizes from $1$ to $4$ Mpc. We look for haloes with at least two progenitors of which at least one has  mass $M\ge M_{cr}(z)$, where $M_{cr}(z)$ is the SV06 critical mass for HD over-cooling. We find that the fraction of over-cooled haloes with mass between  $M_{cr}(z)$ and  $10^{0.2}M_{cr}(z)$, roughly below the atomic cooling limit, can be as high as $\sim0.6$ at $z\approx10$ depending on the merger mass ratio. This fraction decreases at higher redshift reaching a value $\sim0.2$ at $z\approx15$. For higher  masses, i.e. above $10^{0.2}M_{cr}(z)$ up to $10^{0.6}M_{cr}(z)$, above the atomic cooling limit, this fraction rises to values $\ga0.8$ until $z\approx12.5$. As a consequence, a non negligible fraction of high redshift $z\ga10$ mini-haloes can drop their gas temperature to the Cosmic Microwave Background temperature limit allowing the formation of low mass stars in primordial environments.   
\end{abstract}


\section{Introduction}

\begin{table*}
\caption{Simulation parameters. We have adopted $h= 0.719$ thus lengths are in Mpc and masses in $M_{\odot}$.}
\centering
\begin{tabular}{c c c c c c c}
\hline\hline
              &                 &          &               &                  &                                                \\
Sim. Name     & Number of sims. & Box size & Part. number  & Part. mass       & $M_{cr}(z=10)/m_p$ & $M_{cr,1}(z=17.5)/m_p$\\
              &    $N$          & $L_{\rm box}$/Mpc  & $N_p$         & $m_p/M_\odot$    & $N_{p,h}$            & $N_{p,h}$               \\
              &                 &          &               &                  &                      &                         \\
\hline 
              &                 &          &               &                  &                      &                         \\
S1Mpc256      & 20              & 1        & 256$^3$       & 5.88$\times10^3$ & 3.72$\times10^3$     &  1.31$\times10^3$       \\
S2Mpc256      & 20              & 2        & 256$^3$       & 4.70$\times10^4$ & 4.66$\times10^2$     &  1.64$\times10^2$       \\
S4Mpc256      & 20              & 4        & 256$^3$       & 3.77$\times10^5$ & 5.80$\times10^1$     &  2.10$\times10^1$       \\
S1Mpc512      & 5               & 1        & 512$^3$       & 7.35$\times10^2$ & 2.98$\times10^4$     &  1.05$\times10^4$       \\
S2Mpc512      & 5               & 2        & 512$^3$       & 5.88$\times10^3$ & 3.72$\times10^3$     &  1.31$\times10^3$       \\
S4Mpc512      & 5               & 4        & 512$^3$       & 4.70$\times10^4$ & 4.66$\times10^2$     &  1.64$\times10^2$       \\
\hline
\end{tabular}
\label{table1}
\end{table*}

In the current $\Lambda$ Cold Dark Matter ($\Lambda$CDM) cosmological paradigm, dark matter (DM) over-densities are the building blocks of cosmic structures. These DM over-densities grow due to gravity forming DM haloes in a hierarchical way, i.e. from the smaller to the bigger ones, and mergers play an important role in this process. 

For the formation of the first luminous objects to become possible, the baryonic content of the haloes must be able to cool. Cooling of primordial gas is driven by molecular Hydrogen (${\rm H}_2$) which can form  inside DM mini-haloes of  mass $\gtrsim10^6$M$_\odot$. Once ${\rm H}_2$ formation is triggered,  rovibrational transitions of the ${\rm H}_2$ molecule are able to cool the primordial gas down to temperatures of  $\sim$ 200 K \citep{Haiman1996,Tegmarketal1997,Abeletal2002}, see also the  \citet{BarkanaLoeb2001} review. At lower temperatures, the ${\rm H}_2$ lines  become insufficient to cool the gas further.

The ${\rm H}_2$ cooling temperature floor (T$\approx 200$K) and its saturation number density ($n\approx 10^4$cm$^{-3}$ i.e. the density for Local Thermal Equilibrium at which ${\rm H}_2$ cooling is inefficient)  yield a Jeans mass:
\begin{equation}
M_J\approx 500 {\rm M}_\odot \left(\frac{T}{200 {\rm K}}\right)^{3/2} \left(\frac{10^4 {\rm cm}^{-3}}{n}\right)^{1/2}.
\end{equation}
This  sets a mass scale for gravitationally bounded objects in  the primordial gas, suggesting that the first stars  were massive \footnote{But see  \citet{Greifetal2011} and \citet{StacyBromm2013} for  lower masses primordial stellar binary-multiple systems.}.

However, if  the HD molecule is formed in a significant amount in primordial environments, although it has no relevant role in the first stage of ${\rm H}_2$ driven cooling  \citep{Lepp&Shull1983,Brommetal2002} it could  eventually  ($T\lesssim 150$K) become important  \citep{BougleouxGalli1997,Machidaetal2005}, allowing the gas to reach temperatures as low as the Cosmic Microwave Background (CMB) temperature limit at the corresponding redshift. 
If  HD cooling could be triggered, a lower temperature floor for the gas  would decrease the Jeans mass and thus  this process could favor the formation of low mass stars at high redshift.

 

Because the HD number density depends on the H$_2$ abundance through H$_2$ + D$^+\rightarrow$ HD + H$^+$ \citep{Pallaetal1995,GalliPalla2002} and the H$_2$ abundance depends on the free electron number density through e$^-$ + H $\rightarrow$ H$^-$ + $\gamma$ followed by H$^-$ + H $\rightarrow$ H$_2$ + e$^-$ \citep{Peebles1968}, if the gas presents a high ionization fraction it is possible to increase the HD abundance. It has been shown that such high ionization fraction conditions are common in post-shocked gas inside DM haloes \citep{Greifetal2008,jpp3}.
In fact, the DM halo growing process involves violent merger events. These mergers are able to produce strong shock waves which both compress the halo baryonic content and increase the ionization fraction. This drives an enhancement in the formation rate of HD molecules with the consequent over-cooling of the primordial gas, as shown in \citet{Greifetal2008} and \citet{jpp3}. 

\citet{ShchekinovVasiliev2006} (hereafter SV06) studied the necessary  (thermo-chemical) conditions for  HD cooling to switch on.  They argue that such conditions are fulfilled in merging  DM haloes with a total system mass above a critical value, so suitable shocks form. The post-shocked gas with an enhanced HD molecular fraction is able to drop its temperature to the CMB floor of $T_{\rm CMB}\approx 2.73(1+z)$.

SV06 however only considered a straw-man head-on collision of two primordial clouds of equal mass, but, clearly the physical state of the post-shock gas depends on many factors that are not captured by this simplified scenario.
\citet{jpp3} produced numerical DM + baryons simulations  that capture enough physics to study the  physical state of the post-shock gas. They find that as a result of the hierarchical merging process, turbulence is generated and  the production of coolants is enhanced,  so much that even the HD molecule becomes an important coolant in some regions. Yet, their simulations are not sufficient to assess how generic this is, that is, how  these regions   are associated to the distribution of minihaloes. This is what we set up to do here. 


 
In this paper, we use a set of DM cosmological simulations to compute the fraction of haloes able to produce over-cooling of the primordial gas due to mergers at high redshift as predicted by SV06 using the recipe developed in \citet{jpp3}. The paper is organized as follows. In \S 2 we describe our methodology. In \S 3 we show our numerical results and discuss about them. In \S 4 we present our summary and conclusions.

\section{Methodology}
\label{Methodology}

In principle, to compute the fraction of haloes able to over-cool their baryonic content due to mergers at high  redshifts, we would want to have  multiple hydrodynamic simulations, which model both dark matter and baryonic physics  and chemo-thermal evolution of primordial gas,  for cosmological initial conditions, reaching a resolution of  $\sim 1$pc at $z=10$. This  ambitious goal was achieved in  \citet{jpp3} but only for  a single 1 Mpc size box  and in there  the formation and baryonic matter accretion process  of a single halo was  simulated at full resolution: a region of 2 kpc  (at $z=10$)  with $\sim 2$ pc resolution (at $z=10$).   The average CPU-time for one of such systems is $\sim 180000$ CPU-hrs. This   makes it  computationally very expensive  to replicate  the \citet{jpp3} runs for multiple haloes in a cosmological context.  However  this complex problem can be broken in three ingredients  which can be studied independently. The first ingredient is the  thermo-chemical conditions for the HD cooling to switch on which were studied in SV06.   The second ingredient is the physical conditions of the primordial gas (turbulence and shocks) which was studied in \citet{jpp3}. They find that  post shock regions   are able to produce both ${\rm H}_2$ and HD molecules very efficiently even in small mini-haloes ($M\sim 10^6 M_{\odot}$) if they  accrete on, or merge  with,  a more massive but still relatively low mass halo ($M\sim 10^7 M_{\odot}\simeq M_{cr}$). The remaining ingredient is how frequently this happens in   a cosmological context. This last step, however, can be addressed with DM-only simulations under minimal assumptions, and this is what we set up to do here.



\begin{table}
\caption{Relevant redshift and mass bins. The mass  bins labeled by $i=1,2,3$ have been chosen so that bin mass lower and upper boundaries are  $10^{0.2(i-1)}M_{cr}(z_2)$ and $10^{0.2(i)}M_{cr}(z_2)$  respectively. The mass range spanned by the three bins covers the transition from H$_2$ cooling haloes to atomic cooling ones.}
\centering
\begin{tabular}{c c c c c}
\hline\hline
      &       &                   &                   &                   \\
$z_1$ & $z_2$ & Mass bin 1        & Mass bin 2        & Mass bin 3        \\
      &       & in $10^7 M_\odot$ & in $10^7 M_\odot$ & in $10^7 M_\odot$ \\
\hline 
      &       &                   &                   &                   \\
10.0  & 10.2  & 2.10 - 3.33       & 3.33 - 5.28       & 5.28 - 8.36       \\
11.0  & 11.4  & 1.73 - 2.74       & 2.74 - 4.34       & 4.34 - 6.88       \\    
12.3  & 12.7  & 1.41 - 1.81       & 1.81 - 2.87       & 2.87 - 4.55       \\  
14.9  & 15.4  & 0.99 - 1.57       & 1.57 - 2.49       & 2.49 - 3.94       \\
17.2  & 17.9  & 0.74 - 1.17       & 1.17 - 1.85       & 1.85 - 2.94       \\  
\hline
\end{tabular}
\label{table2}
\end{table}

\begin{table}
\caption{Total number of OC haloes (i.e. the sum over the $N$ realizations) and total number of haloes ($N_h$) in a given mass bin  with $f_{mg}=0.6$ and $m_p\le5.88\times10^3M_{\odot}$ at $z_1=10$.}
\centering
\begin{tabular}{c c c c c}
\hline\hline
          &           Mass  &       Mass     &   Mass    & Total    \\
Sim. Name &   bin 1 &  bin 2 & bin 3 &  Volume  \\
          &  $N_{OC}(N_{\rm h})$   & $N_{OC}(N_{\rm h})$   & $N_{OC}(N_{\rm h}) $ & Mpc$^3$ \\
\hline 
          &             &            &         &   \\
S1Mpc512  & 5 (7)          & 4 (4)          & 3 (3)       & 5  \\
S1Mpc256  & 25 (46)        & 14 (15)        & 11 (12)     & 20   \\    
S2Mpc512  & 88 (132)         & 56 (57)         & 37 (38)      &  40 \\  
\hline
\end{tabular}
\label{table3}
\end{table}

The critical mass to trigger HD molecular over-cooling at redshift $z$ is defined by SV06 as:
\begin{equation}
M_{cr}^{SV06}(z)=8\times10^6\left(\frac{20}{1+z}\right)^2 M_{\odot}.
\end{equation}
Following SV06 this is  the total mass of the system, i.e. DM plus baryonic mass. Because we have to work  with   DM only  simulations, we have to make an assumption  regarding  the baryonic matter. To take into account the gas inside the DM haloes we assume that these primordial  haloes host the universal baryon fraction
\begin{equation}
\frac{M_b}{M_{DM}+M_{b}}=\frac{\Omega_b}{\Omega_m}\equiv f_b,
\end{equation}
where $M_b$, $M_{DM}$, $\Omega_b$, $\Omega_m$ and $f_{b}$ are the baryonic mass content of the halo, the dark mass content of the halo, the current average baryonic matter density in the Universe in units of the critical density, the current average DM density in the Universe in units of the critical density and the universal baryonic mass fraction of the Universe, respectively. 
Using this approximation, the necessary (but not  yet sufficient) condition for a DM halo  at redshift $z$ to become an over-cooled halo is that it must have a  DM mass above a critical mass  
\begin{equation}
M_{cr}^{DM}(z)=(1-f_B)\times M_{cr}^{SV06}(z),
\end{equation}
hereafter we will refer to $M_{cr}^{DM}$ as $M_{cr}$.


We use the cosmological hydrodynamical code \texttt{RAMSES} \citep{Teyssier2002} to perform $75$ DM-only simulations. The cosmological initial conditions are produced with the \texttt{mpgrafic} code \citep{Prunetetal2008} and the initial redshift for each run is set to $z_i\approx65$. The cosmological parameters  are those of the concordance $\Lambda$CDM model   from \citet{Komatsu2009,Komatsu2010}: $\Omega_m=0.258$, $\Omega_\Lambda=0.742$  $h=0.719$, $\sigma_8=0.796$, $n_s=0.963$ and the transfer function of \citet{EisensteinHu1998} with  $\Omega_b=0.0441$.

Using the \texttt{AHF} halo finder \citep{AHF2009}, we identified DM haloes (i.e., objects with a density contrast $\delta\ge 200$) with mass above or equal to the critical mass to enhance the HD molecular cooling $M_{cr}$, at several  redshifts $z\ge10$.  For reference, in the cosmology adopted here,  $f_b=0.1709$,  and thus 
\begin{equation}
M_{cr}(z)=6.63\times10^6\left(\frac{20}{1+z}\right)^2 M_{\odot}\,.
\end{equation}
In the set-up chosen for the \texttt{AHF} halo finder
the minimum particle number per halo was set to $N_{min}=20$. We adopted this low number because we are not interested in characterizing the haloes based on their internal-radial features. This corresponds closely to the minimum number of particles per critical mass halo  $N_{p,h}$ at the highest  redshift of interest in the lower resolution run.

Table \ref{table1} shows the details of each simulation. From the first column to the last one: the simulation name, referring to both the box size and the particle number, the number of simulations $N$, the box size\footnote{Note that we adopt the value $h=0.719$ therefore here length are in Mpc and masses in $M_{\odot}$.} $L_{\rm box}$ in Mpc , the number of particles per simulation $N_p$, the  particle mass $m_p$ and the number of particle per critical mass halo $N_{p,h}$ at two reference redshift $z=10$ and $z=17.5$.

\begin{figure}
\centering
\includegraphics[height=6cm]{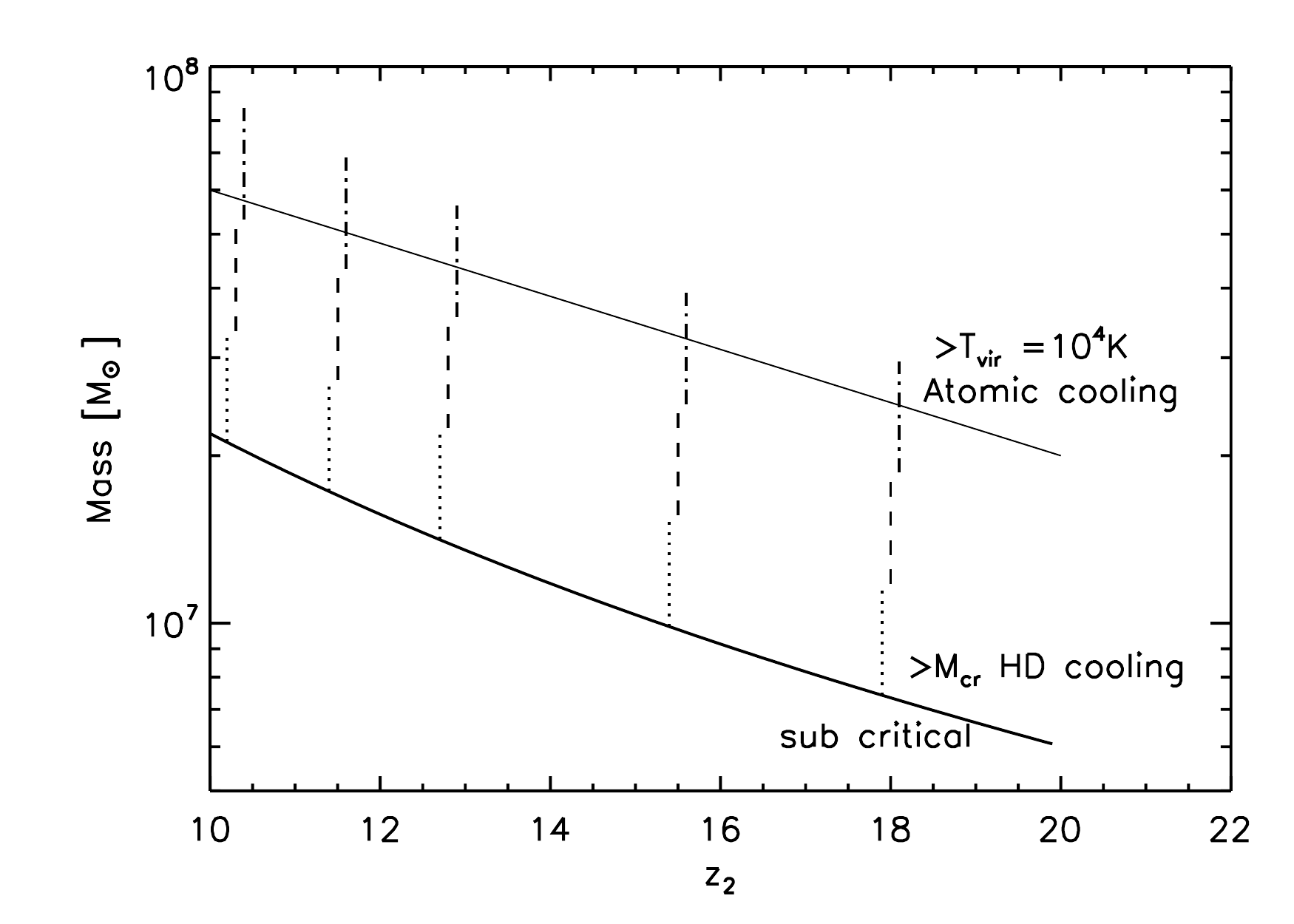}
\caption{The mass scales involved. The lower solid line shows the critical mass for HD cooling $M_{cr}(z)$, the upper solid line corresponds to $T_{\rm vir}=10^4$K necessary for atomic lines cooling. The vertical bars (dotted, dashed and dot-dashed) correspond to the three mass bins considered.}
\label{fig:mass}
\end{figure}

\begin{figure*}
\centering
\includegraphics[height=12cm,width=16cm]{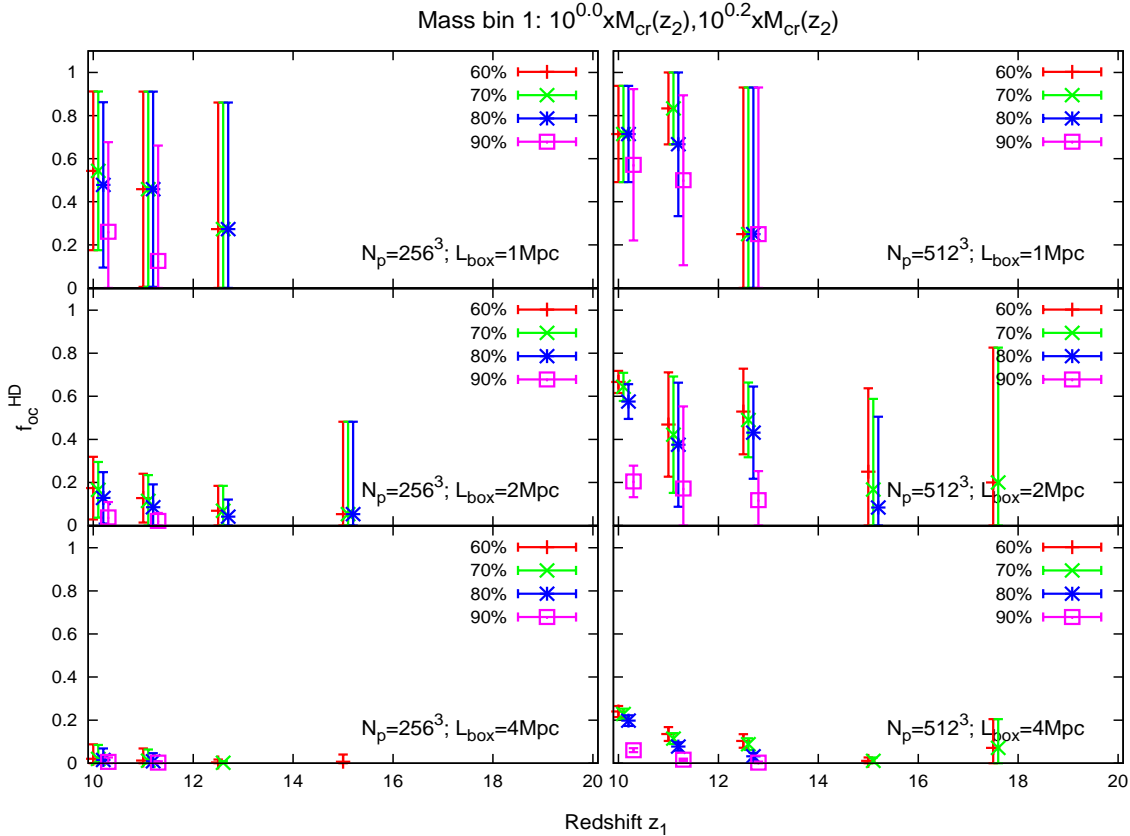}
\caption{Fraction of over-cooled haloes $f_{\rm OC}^{\rm HD}$ in the first mass bin centered around $M=1.3M_{cr}(z_2)$ (see table \ref{table2}) for different merger parameters $f_{mg}$=0.6,0.7,0.8 and 0.9, indicated by  plus , 'x',  stars and square symbols respectively. The results are shown  for each redshift $z_1$ from table \ref{table2}: $z_1=$10.0, 11.0, 12.3, 14.9 and 17.2. The left column shows the low resolution runs with $256^3$ particles and the right column shows the high resolution runs with $512^3$ particles. The upper row shows the 1Mpc runs, the middle row shows the 2Mpc runs and the bottom row shows the 4Mpc runs. The high-resolution, numerically converged set-up corresponds to the top two panels and the middle row right panel.}
\label{fig1}
\end{figure*}

\begin{figure*}
\centering
\includegraphics[height=12cm,width=16cm]{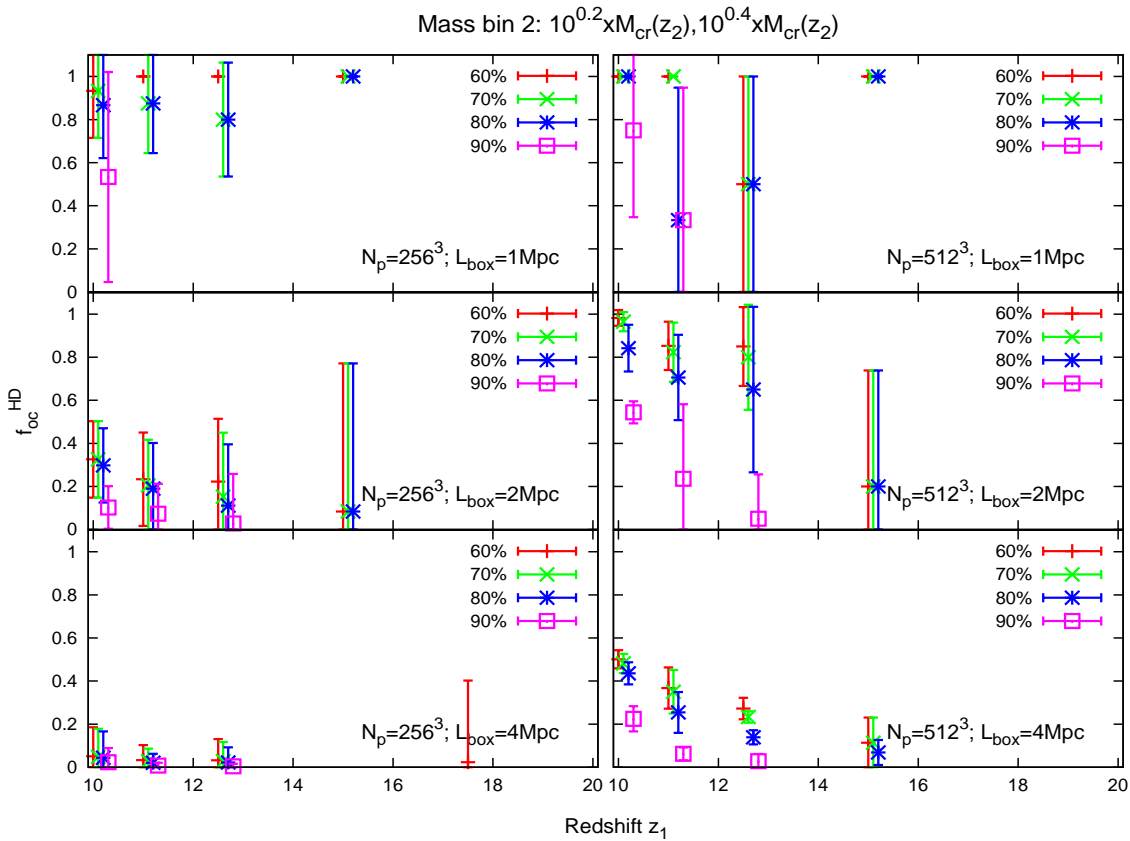}
\caption{Same as figure \ref{fig1} but for the second mass bin centered around $M= 2 M_{cr}(z_2)$ . See table \ref{table2}.}
\label{fig2}
\end{figure*}

\begin{figure*}
\centering
\includegraphics[height=12cm,width=16cm]{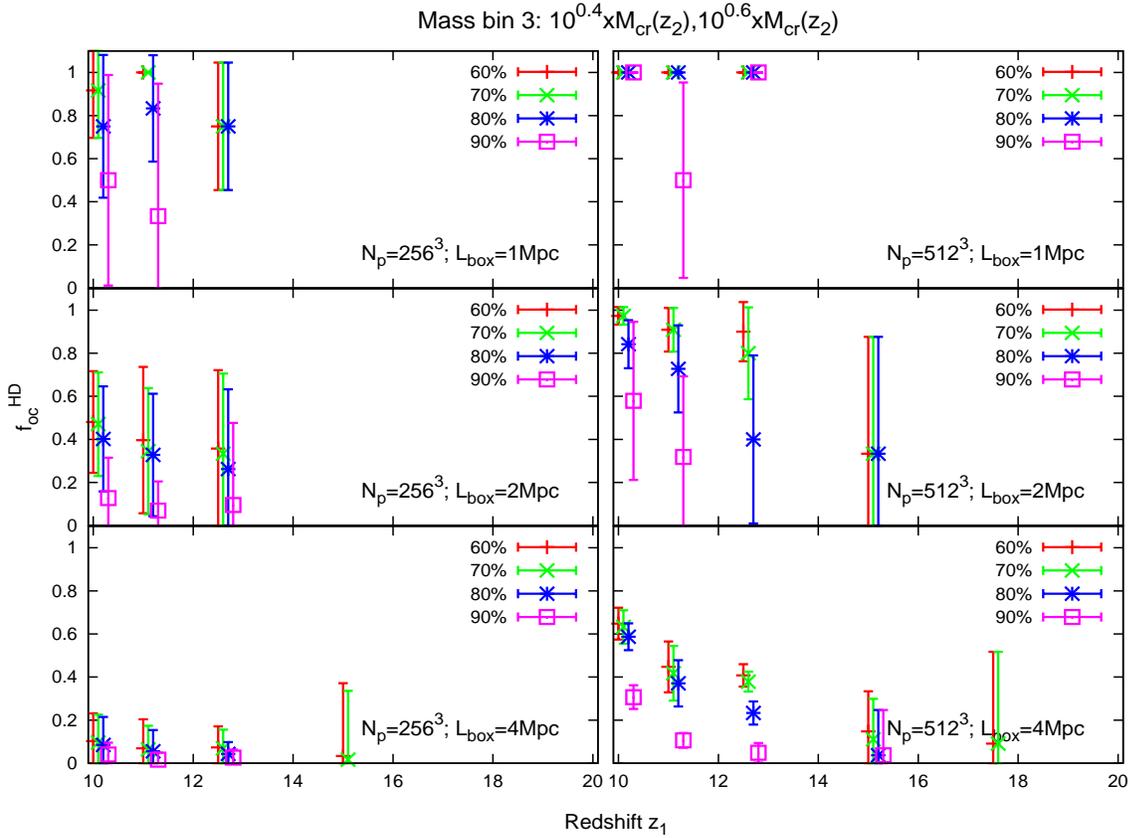}
\caption{Same as figure \ref{fig1} but for the third mass bin centered around $M= 3.2 M_{cr}(z_2)$. See table \ref{table2}.}
\label{fig3}
\end{figure*}

 \begin{figure*}
\centering
\includegraphics[height=8cm,width=12cm]{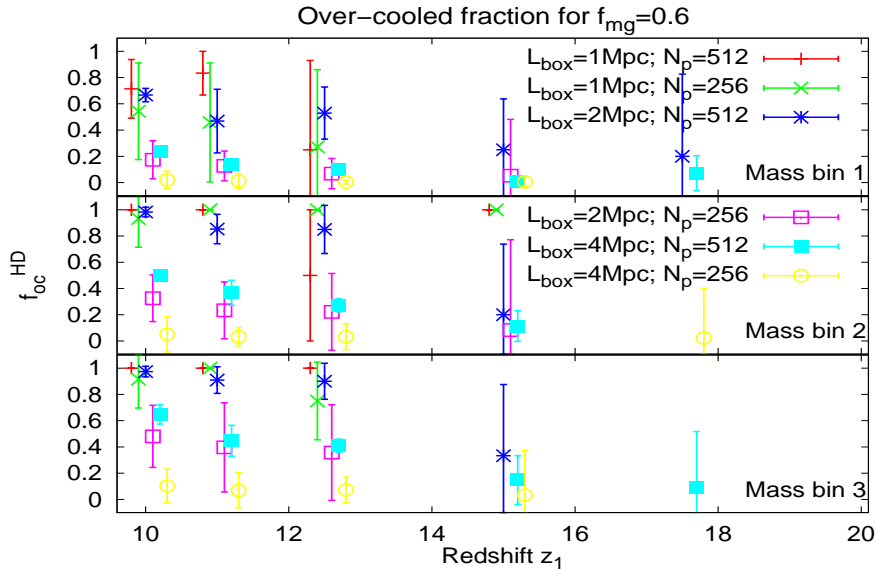}
\caption{Over-cooled fraction for the less restrictive case $f_{mg}=0.6$ as a function of redshift. Different symbols correspond to  our six different simulations: (red) "plus" symbol for S1Mpc512 ,  (green) ``x" symbols for S1Mpc256 , (blue) asterisks for S2Mpc512 ,(magenta) empty squares for S2Mpc256, (cyan) filled squares for S4Mpc512 and  (yellow) circles for S4Mpc256. 
A mass-resolution dependent trend appears:  S1Mpc512, S1Mpc256 and S2Mpc512  with with $m_p\le5.88\times10^3M_{\odot}$ have consistent $f_{\rm OC}^{\rm HD}$. 
S2Mpc256, S4Mpc512 and S4Mpc256 with  $m_p\ge4.70\times10^4M_{\odot}$ have  resolution-dependent  $f_{\rm OC}^{\rm HD}$ with is lower but increases with resolution.}
\label{fig4}
\end{figure*}

\begin{figure*}
\centering
\includegraphics[width=2\columnwidth,height=16cm]{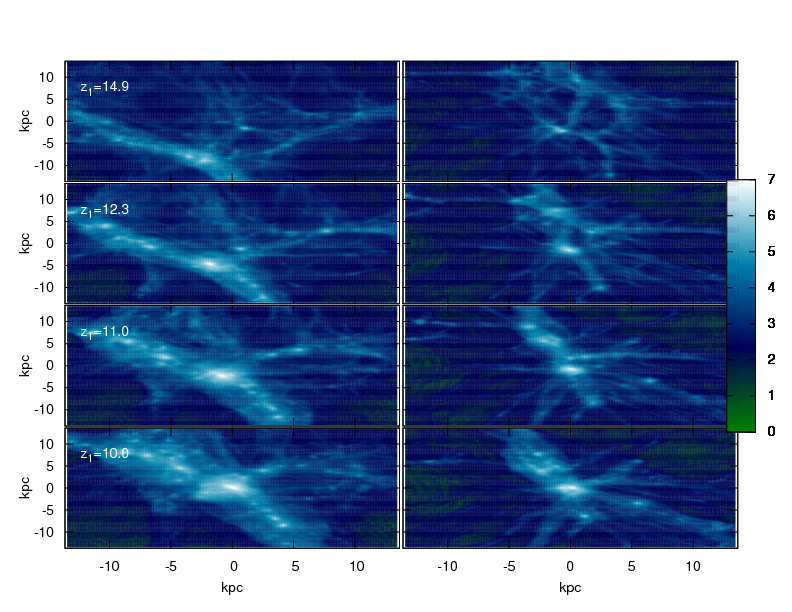}
\caption{Dark matter mass projection in arbitrary units for two randomly chosen OC haloes. The halo on the left  belongs to mass bin 3 ($M=6.41\times10^7M_\odot$ at $z=10$) and the halo on the right   to mass bin 1 ($M=2.53\times10^7M_\odot$  at $z=10$). Each row corresponds to a different redshift $z_1$. The distance scales are in co-moving kpc.}
\label{fig5}
\end{figure*}

As we will show in the next section, the most reliable results come from runs where $M_{cr}$ is defined by $N_{p,h}\ge1310$ particles at $z\le17.5$, i.e. runs with a particle mass $m_p\le5.88\times10^3M_{\odot}$.  In these runs the DM haloes are found consistently in successive snapshots. Furthermore, it is worthwile to note that in these reliable runs the primordial perturbation distance scale $\lambda_{M_{cr}}$ associated to the critical mass $M_{cr}$ is well defined by a number of partcles ($>10$) when the simulation start.  



The \citet{jpp3} findings indicate that a necessary and sufficient condition for triggering HD cooling is that  a halo with mass  greater than $M_{cr}$ (recall that at the redshifts of interest  $M_{cr} \sim 10^7 M_{\odot}$) undergoes a merger or accretes baryonic material funnelled in the halo along  filaments. Even in sub-critical haloes  ($M\sim10^6 M_{\odot}$) HD cooling can be triggered if they accrete on a critical one, as the relevant physical condition driving the turbulence is the relative velocity, which is set by the  potential well created by the super-critical halo.  In the super-critical halo, if it is not  disrupted by a major merger, the turbulence triggered by accretion is enough enhance the creation of H$_2$ and HD and therefore kickstart over-cooling.

Informed by the above findings, here we impose the conditions for over-cooling to happen as follows.

We construct the merger trees for each simulation using the \texttt{AHF} merger tree tool, we identify  DM haloes  at redshift $z_2$ with mass  $\geq M_{cr}(z_2)$ that subsequently undergo  merging to form a bigger halo at $z_1$ (with $z_2\ge z_1$).


We define the over-cooled (OC) halo merger as the process in which an existing halo at $z_1$ has at least two progenitors,  of which at least one with mass $M_{DM}\ge M_{cr}(z_2)$ and after the merger  keeps at least a mass fraction $f_{mg}$ of the most massive progenitor.
We vary the factor $f_{mg}$ from 0.6 to 0.9 in order to study how the OC haloes fraction depends on it.


Our parameters of the halo finder routine imply that  the minimum halo mass (which sets therefore the definition of merger) is somewhat resolution-dependent  ranging from $1.47\times 10^4 M_{\odot}$ in simulation S1Mpc512  to $7.54\times10^6$ in  simulation S4Mpc256.

Here we want to stress that  we do not impose a minimum merger mass ratio  in our strategy to look for OC haloes. The method described above is suitable to address the question: {\it what is the fraction of DM haloes  able to over-cool their baryonic content (and thus potential site for low mass star formation) due to mergers and accretion at high redshift?}. Indeed, the condition $M\ge M_{cr}$ ensures that the interaction between haloes will be strong enough to trigger the enhancement of the HD formation. On the other hand, a study based on the merger mass ratio could give us information about the amount of OC gas and then could help  answer a different question: {\it what is the amount of OC gas in haloes at high redshift?}.  Our simulation set up and our  methodology cannot quantify  the amount of over-cooled gas but it is suitable to estimate the fraction of OC haloes at high redshift. This is the goal of the present work. 

%

\section{Results and Discussion}

Table \ref{table2} shows some  example combinations of redshifts $z_1$ and $z_2$ used in building our merger tree, and the ranges of the three  halo mass bins (at $z_2$)  we  consider. The mass  bins labeled by $i=1,2,3$ have been chosen so that bin mass lower and upper boundaries are  $10^{0.2(i-1)}M_{cr}(z_2)$ and $10^{0.2(i)}M_{cr}(z_2)$  respectively. Thus the three mass bins are centered around, 1.3, 2.0, and 3.2 $M_{cr}(z_2)$, respectively.  With this choice, the mass range spanned by the three bins covers the transition from H$_2$ cooling mini-haloes (with virial temperature $T_{vir}\la$few$\times10^3$K and $M_{vir}\la 1.5\times10^7 M_\odot$) to atomic cooling haloes (with $T_{vir}\ga\times10^4$K and $M_{vir}\ga 1.5\times 10^7 M_\odot$). This is summarized in Fig.\ref{fig:mass}.

Table \ref{table3} reports  the total  number of OC haloes $N_{OC}$ and the total number of haloes $N_{\rm h}$ at $z_1=10$ for the three different mass bins in the less restrictive case $f_{mg}=0.6$ and for the highest resolution simulations with $m_p\le5.88\times10^3M_{\odot}$. The reported number is the sum of all OC haloes  (all haloes) in the  $N$ simulations considered, i.e.,  5 simulations of S1Mpc512, 20 for S1Mpc256 and 5 for S2Mpc512.  The effective  volume for finding these OC haloes is therefore 5 Mpc$^3$, 20 Mpc$^3$ and 40 Mpc$^3$, respectively.

%

Figures \ref{fig1}, \ref{fig2} and \ref{fig3} show the fraction of OC haloes $f_{\rm OC}^{\rm HD}$ for four different values of $f_{mg}$ and for the three halo mass bins as a function of redshift. These results are shown for our two different $N_p$ (in different columns) and  different box sizes $L_{\rm box}$ (in different rows). The error bars correspond to the standard deviation  between the $N$ simulations  at a given redshift.  The error on the mean would be smaller by a factor $\sqrt{N}$.

As we expected, the higher the $f_{mg}$ the lower the OC fraction $f_{\rm OC}^{\rm HD}$. This trend shows that after a merger process it is very difficult for the resulting halo to keep  100$\%$ of its progenitor's mass: some of the  progenitor's mass  is always removed from the parent halo after the merger.  The resulting $f_{\rm OD}^{\rm HD}$ shows a very weak dependence (or no dependence at all)  on $f_{mg}$ for $0.6\leq f_{mg}\leq 0.8$.


Our results show a clear resolution dependence for $m_p\ge4.70\times10^4M_{\odot}$, i.e. runs S4Mpc512, S4Mpc256 and S2Mpc256. In these runs it is possible to see a monotonic growth of the OC fraction with the simulation resolution, which is particularly marked in the $f_{mg}=0.6$ case: we thus discuss numerical convergence before further interpreting Figs.~\ref{fig1}, \ref{fig2} and \ref{fig3}.

Numerical convergence is  investigated further in Fig.~\ref{fig4}  where it is possible to identify a mass-resolution dependent trend. Simulations S1Mpc512,  S2Mpc512  (and S1Mpc256)  have  particle masses below  $m_p= 5.88\times10^3M_{\odot}$  and thus a mass threshold for OC halo merger $M=1.2 \times 10^5$ ($M=1.47\times10^4$) or mass merger ratios below $1:65$. These simulations correspond to the (red) plus symbols, (blue) asterisk symbols and (green) ``x'' symbols.  At this resolution results for $f_{\rm OC}^{\rm HD}$  appear to converge. On the other hand  simulations  S2Mpc256, S4Mpc512 and S4Mpc256 with $m_p\ge4.70\times10^4M_{\odot}$ , (magenta) open square symbols, (cyan) filled square symbols and (yellow) open circle symbols, do not show numerical convergence. This can be understood as the mass threshold for merger in these simulations is high ($> 9.4\times 10^5 M_{\odot}$) and the merger mass ratios are larger than $1:2$.

%
%


 In what follows we will focus on these three higher mass resolution   simulations because they have the most reliable results based on both, convergence and number of particles per DM halo.

In figure \ref{fig1}, corresponding to the first mass bin (see table \ref{table2}), and for mass resolution $m_p\le5.88\times10^3M_{\odot}$ (i.e., top two panels,  and middle right panel) our results show that at $z=10$ the fraction of OC haloes is $f_{\rm OC}^{\rm HD}\ga0.5$ in  the case with $f_{mg} \lesssim  0.7$. This fraction tends to decrease at higher redshift ($z\la12.5$) but is always  above  20$\%$ (for $f_{mg}\lesssim 0.7$) showing that a non negligible fraction of DM haloes in this mass bin is able to over-cool their gas content due to mergers at high redshift. At higher redshifts, i.e., $z\ga15$, the fraction decreases $f_{\rm OC}^{\rm HD}\la0.2$. This last result comes from S2Mpc512, the only simulation with data above $z\ga15$ in this mass bin.

In figure \ref{fig2}  we show the  second mass bin centered around $M=2 M_{cr}(z_2)$. For runs with mass resolution $m_p\le5.88\times10^3M_{\odot}$ (top two panels,  and middle right panel) the OC fraction at $z=10$ is $f_{\rm OC}^{\rm HD}\ga0.9$ for  $f_{mg}\lesssim 0.7$ and it can reach $f_{\rm OC}^{\rm HD}\sim 1.0$. At higher redshift ($z\la12.5$) the fraction remains significant,  $f_{\rm OC}^{\rm HD}\ga0.8$. As expected the OC fraction increases with the mass of the halo.

Figure \ref{fig3} shows our results for the third mass bin centered around $M=3.2 M_{cr}(z_2)$. The OC fraction keeps increasing with halo mass.

In summary  these figures show that  a non negligible fraction of DM haloes  above the critical mass $M_{cr}$ are able to over-cool their gas content due to mergers at high redshift.

To illustrate  how the OC merger proceeds, figure \ref{fig5} shows the evolution of two randomly chosen OC haloes from our catalog at 4 different redshift $z_1$. In the first column   we show  an OC halo of $M=6.41\times10^7M_\odot$ (computed at $z=10$) from the third bin mass and in the second column  a $M= 2.53\times10^7M_\odot$ (computed at $z=10$) OC halo from the first  mass bin. The difference in size of the objects reflects the different mass bins.

As an additional study, we have computed the probability distribution function for the halo spin parameter $\lambda$ defined by \citet{Bullock2001} and we have found that it follows  log-normal distribution characterized by a standard deviation $\sigma\approx0.5$ and an average spin parameter $\bar\lambda\approx0.04$ in good agreement with previous works, e.g. \citet{DavisNatarajan2009}. Despite of the low number of haloes in the most reliable runs (see table \ref{table3}), we recover a log-normal distribution (for S1Mpc256 and S2Mpc512) characterized with the parameters shown above. This fact supports our claim on the reliability of our results.


While the N-body simulations for this work were running, new results on cosmological parameters, derived from the {\it Planck} satellite observations,  were released \citep{PlanckCollaboration2013}.  The {\it Planck}'s best fit $\Lambda$CDM cosmological parameters are somewhat different from WMAP's ones. Because our results can be cosmology-dependent, let us  elaborate on the possible  effect of the {\it Planck} results. {\it Planck}'s $\Omega_m$ value is slightly higher than WMAP's and  $\Omega_b$  slightly lower. This affects directly the computation of the  DM critical mass $M_{cr}(z)$ decreasing it by a $2\%$, approximately. Furthermore, because the {\it Planck's} value of the Hubble constant is lower, each redshift in our calculations has to be increased by about  $4\%$ and  so the box size $L_{\rm box}$. Thus  the  changes associated to the new best fit  cosmological parameters have a negligible effect on our results.
%
 
\section{Summary and Conclusions}

We have performed $75$ DM-only cosmological simulations with two different particle numbers ($256^3$ and $512^3$) and inside three different box sizes ($L_{\rm box}=$ 1Mpc, 2Mpc and 4Mpc) in order to quantify the fraction of haloes able to over-cool their baryonic content due to mergers at high redshift as predicted by \citet{ShchekinovVasiliev2006}. 
 
 As shown in \citet{jpp3} accretion and  (minor) mergers onto a halo of mass above the  critical value   defined by SV06, $M_{cr}(z)$ produce supersonic turbulence and a shocked environment where H$_2$ and HD molecules are formed efficiently. There, regions are able to (over)cool below the H$_2$ cooling temperature floor.
 
To identify the fraction of haloes where the above conditions are verified,  we computed the progenitor's mass for each halo at a given redshift inside a bin mass, specified in table \ref{table2}. This mass range  spans the transition between H$_2$ molecular cooling  to atomic cooling haloes. Every halo with more than one progenitor  of which at least one has a  mass above  $M_{cr}(z)$, was counted as an over-cooled (OC) halo.

Our results show that a non negligible fraction of the mini-haloes formed at $z\ge10$ over-cooled their primordial gas due to the process outlined above. The fraction of OC haloes at $z=10$ is $f_{\rm OC}^{\rm HD}\ga0.5$   for masses  roughly below the atomic cooling limit: $1\times10^7\la M/M_\odot\la3\times10^7$. At higher redshift, $z\la12.5$, the fraction $f_{\rm OC}^{\rm HD}\ga0.2$ and it is below 0.2 for $z\ga15$. The fraction of OC haloes rises with halo mass. For haloes above the atomic cooling limit, $2\times10^7\la M/M_\odot\la8\times10^7$,  the fraction of OC haloes at $z\la12.5$ is $f_{\rm OC}^{\rm HD}\ga0.8$. 

The  existence of a non negligible fraction of OC haloes at high redshift has interesting consequences for the star formation process in primordial environments. As predicted by SV06 the HD molecular cooling drops the gas temperature to the CMB limit $T_{\rm CMB}(z)\approx2.73(1+z)$ allowing the formation of low mass primordial stars \citep{jpp2,jpp3}.  Their low mass makes these primordial (population III)  stars  very long-lived opening  a window for the potential detection of primordial stars in the local Universe. 

\section*{Acknowledgements}

JP thanks Roberto Gonzalez and Christian Wagner for their useful and constructive comments on this work. JP, RJ and LV  acknowledge support  by  Mineco grant FPA2011-29678- C02-02.  LV is supported by European Research Council under the European CommunityÕs Seventh Framework Programme grant FP7- IDEAS-Phys.LSS.

\end{document}